
\input harvmac
\lref\swh{S.W. Hawking, {\sl Commun. Math. Phys.} {\bf 43} (1975) 199.}
\lref\bek{J.D. Bekenstein, {\sl Phys Rev.} {\bf D7} (1973) 2333;
{\sl Phys Rev.} {\bf D9} (1974) 3292.}
\lref\suss{J.G. Russo and L. Susskind, hep-th/9405117.}
\lref\sen{A. Sen, hep-th/9504147.}
\lref\DS{M. Dine and N. Seiberg, {\sl Phys. Rev. Lett.}
{\bf 55} (1985) 366.}
\lref\gv{J. Garriga and A. Vilenkin, {\sl Phys Rev.} {\bf D47}
(1993) 3265.}
\lref\gva{J. Garriga and A. Vilenkin, {\sl Phys. Rev.} {\bf D44} (1991)
1007.}
\Title{\vbox{\baselineskip12pt\hbox{SUSX-TH-95/33}\hbox{gr-qc/9508031}}}
{How is a Closed String Loop like a Black Hole?}
\centerline{E. J.
Copeland\footnote{$^{\dag}$}{E.J.Copeland@central.susx.ac.uk}
and Amitabha Lahiri\footnote{$^{\ddag}$}{A.Lahiri@central.susx.ac.uk}}
\bigskip\centerline{\it School of Mathematical \& Physical Sciences}
\centerline{University of Sussex}
\centerline{Falmer, Brighton, BN1 9QH, UK}

\def\lq{\log{\cal Q}}
\def\mb{{\beta M\over 2}}
\def\bh{black hole}

\def\ap{\alpha'}
\def\schw{Schwarzschild\ }
\def\vev#1{\langle #1 \rangle}
\def\oz{\omega_0}
\def\ibh{\int^{\omega_0}_{1/R}}
\def\dmdb{{dM\over d\beta}}
\bigskip\centerline{\bf Abstract}
We demonstrate that under plausible assumptions the entropy and
temperature associated with the small oscillations on a
circular loop of radius $R$ and a black hole of mass $M=R/2G$ are
identical.
\Date{5/95}

According to Hawking \swh\  and Bekenstein \bek,
the inverse temperature and entropy of
a \schw \bh\ of mass $M$ are respectively given by
\eqn\htemp{\beta_{bh} = {1\over T} = {8\pi GM\over \hbar},\qquad
S_{bh} = {4\pi GM^2\over \hbar } = \mb.}
If we consider $E_{bh} = M$ as the energy of this
system, then the free energy is given by
\eqn\free{F = E_{bh} - T_{bh}S_{bh} = M/2.}
Ever since they were published, these relations have given rise to much
speculation in both gravity and quantum field theory.  Is the evolution of
a quantum state non-unitary in the presence of a black hole? Do ultraviolet
divergences of quantum fields modify gravitational singularities? What has
surface area got to do with entropy?

In this letter we add to the speculation and report a most peculiar
coincidence. The formulae relating the mass, entropy and temperature of a
black hole $including$ the numerical coefficients, are identical to those
of a simple model, namely, a system comprising of small oscillations on a
closed string wrapped around what would be the event horizon of a \schw
\bh\ of the same mass. Although intriguing, this result seems to rely on a
few key assumptions, every one of which can be justified on physical
grounds. We shall point out these assumptions as we go along.


To begin with we consider as our statistical system the small oscillations
on a closed circular string.  In four dimensions there are two
polarizations, $\phi_r$ and $\phi_\perp$, one in the radial direction and
the other perpendicular to it.  Following \gva, we write the coordinates of
the perturbed world-sheet as
\eqn\param{\tilde x^\mu = x^\mu + \sum\limits_{A=r,\perp}n^{A\mu}\phi^A.}
The tranverse mode behaves like a massless
minimally coupled scalar
\eqn\perpeq{\ddot\phi_{\perp} = {1\over R^2} \phi''_{\perp},}
where a dot denotes a time-derivative, prime denotes a $\theta$-derivative,
and $R$ is the radius of the loop at rest (at $t = 0$).  The radial mode
satisfies a more complicated equation, which for a circular loop of string
can be written as \gv
\eqn\radeq{\ddot\phi_r = {1\over R^2}\phi''_r + \cos^2(t/R)K_{ab}K^{ab}\phi_r.}
Here $K_{ab}$ are components of the extrinsic curvature $K_{ab} =
-\del_an^r_\mu\del_bx^\mu$, corresponding to the radial normal vector to
the string worldsheet, $n^r$. The general solution for this equation is
known. Let us consider the perturbations near $t = 0$ (when its radius was
$R$).  In both cases, the solutions are separable, with the Fourier
components labeled by integers $n$ corresponding to the wave number of the
standing waves on the string.  For each polarization state there are two
modes (left and right-moving).  The density of frequencies is therefore
given by
\eqn\density{g(\omega)d\omega \approx\ 4R d\omega.}
where the approximation refers to replacing
$n/\sqrt{n^2 -2}$ by 1 for the radial modes.
In our model, we will look at the partition function due to these
oscillators living on the string loop.

The partition function for an oscillator (massless or massive) of angular
frequency $\omega$ is $z(\omega) = (2\sinh\half\beta\hbar\omega)^{-1}$.
This includes the contribution from the zero-point energy. In standard
calculations of continuum quantum field theory, the zero-point energy is
set to zero by normal ordering on the grounds that only differences in
energies matter in transition processes. On the other hand, here we are
trying to model a gravitational system. The zero-point energy acts as a
source of the gravitational field and therefore cannot be ignored. Also,
the gravitational field provides a natural ultraviolet cutoff which bounds
the partition function from above.

We choose the cutoff $\oz$ by the following semi-classical argument.  The
energy $m_0 = \hbar\oz$ has a corresponding Schwarzschild $diameter$ of
$4G\hbar\oz$. It also has a corresponding Compton wavelength $\lambda_0 =
\hbar/m_0 = 1/\oz$. We equate with the two lengths to obtain a cutoff
\eqn\frcut{\oz = \sqrt{1\over 4G\hbar}.}
The associated energy is the maximum that can be reached by a quantum field
in the sense that a quantum of energy bigger than this will be fully inside
its own event horizon. There is also a natural lower bound on the
frequencies, given by the frequency $\omega_L$ that has a wavelength as
large as the string itself, $\omega_L = 1/R$.

We can now compute the partition function ${\cal Q}$ to be given by
\eqn\pfcalI{\eqalign{\lq &=
-4R\ibh\log(2\sinh\half\beta\hbar\omega)d\omega\hfil\cr &=
-R\beta\hbar(\oz^2 - {c^2\over R^2}) - 4R\ibh\log(1
- e^{-\beta\hbar\omega})d\omega\hfil\cr &= - R\beta\hbar(\oz^2
- {1\over R^2}) - 4R\bigg[\omega\log(1 -
e^{-\beta\hbar\omega})\bigg|^{\oz}_{1/R} -
{1\over\beta\hbar}\int\limits^{\beta\hbar\oz}_{\beta\hbar
/R}{xdx\over e^x - 1}\bigg]. \hfil\cr}}
Now let us look at this in the approximation $\beta\hbar\oz>>1$ (i.e. a
cold system). Then this formula becomes
\eqn\pfcalII{\lq \approx - R\beta\hbar\oz^2 + {\beta\hbar\over R}
+  4\log(1 - e^{-\beta\hbar/R}) + {4R\over\beta\hbar}
\int\limits^{\beta\hbar\oz}_{\beta\hbar/R}{xdx\over e^x - 1}.}
It is easy to verify that in this equation, the last two terms are always
negligible compared to the first term in the limit $\beta\hbar\oz>>1$. (The
integral is finite and smaller in value than $\pi^2/6$.)  Then we have
\eqn\pfcalIII{\lq \approx -{R\beta\hbar}\oz^2 + {\beta\hbar\over R}.}
Let us now replace $R$ by the corresponding \schw mass $M,\ R =
2GM/c^2$, and $\oz$ by its value in \frcut. We then have
\eqn\pf{\lq = -\mb + {2\hbar^2\oz^2\beta\over M}.}
So far, we have not actually made any assumption about what this $M$ means
physically in the context of our system. Here we specify its meaning by
assuming that $M$ is in fact the average energy $\vev E$, in the
thermodynamic sense, of the string. We also assume that $\beta$ and $M$ are
not independent of each other, $i.e.$, one cannot be varied while keeping
the other fixed. Then we have
\eqn\thermI{\vev E \equiv -{d\over d\beta}\lq = {M\over 2} +
{\beta \over 2}\dmdb - {2\hbar^2\oz^2\over M} +
{2\hbar^2\oz^2\beta\over M^2}\dmdb,}
so that using $\vev E = M$ we obtain
\eqn\thermIb{ {M\over 2} +
{2\hbar^2\oz^2\over M} = \dmdb{\beta\over M}\big({M\over 2} +
{2\hbar^2\oz^2\over M}\big).}
Solving this we get the relation between $\beta$ and $M$,
\eqn\thermII{\beta = bM,}
where $b$ is an arbitrary constant. This is a non-trivial result in that we
have obtained a linear relationship between $\beta$ and $M$. A priori there
was no reason for this to emerge. The entropy of this system is then given
by
\eqn\modent{S = \beta\vev E + \lq = b({M^2\over 2} + 2\hbar^2\oz^2).}
%
Apart from the arbitrary constant $b$, we have managed to get an entropy
and an inverse temperature that are tantalizingly similar to the
Hawking-Bekenstein formulae. This system therefore behaves much like a
\bh, getting hotter as it radiates its energy away.

It should be obvious that this result could have been obtained without
assuming any stringy property of our system. The system might have been
scalar fields living on a circle, since we are effectively neglecting the
extrinsic curvature terms in equation \radeq. In particular no reference to
the string tension needs to be made. However it turns out that we can
deduce the value of the proportionality constant $b$ by making an appeal to
string theory, as it introduces a length scale associated with the string
tension. Suppose we think of our system as small oscillations on a closed
bosonic string.  All but four space-time dimensions are then assumed to be
compact, and their scale of compactification are above the cutoff we have
introduced.  Then the previous analysis remains valid, but we have a new
ingredient. As the string radiates energy it gets smaller, and the lower
bound on the frequencies starts climbing up to the upper bound.  Going over
our analysis above, we can see that we run into inconsistencies when the
upper and lower limits of the integral in \pfcalI\ coincide. (The average
energy $\vev E$, as well as $\lq$, vanish, but the radius of the string is
non-zero, which is inconsistent with the model.) In other words, the
canonical analysis of the model fails when
\eqn\fail{{1\over R} = \oz,\ i.e.,\ {\rm when\qquad} M = M_0 = {1\over
2G\oz}.}
On the other hand, using \thermII\ we can say that this failure occurs when
the system gets elevated to a critical temperature. It is tempting to
equate this temperature to the Hagedorn temperature $\beta_H$ of the closed
bosonic string. Although we have no particular reason for choosing it as
opposed to some other temperature slightly lower, the result we obtain is
so compelling so as to make further investigation well worth pursuing. We
also notice that the conserved energy of a closed bosonic string is $E =
R\hbar/\ap$. Equating the statistical conserved energy $\vev E$ to the
dynamical conserved energy $E$, we get an expression for $\ap$. Inserting
that into the expression for the Hagedorn temperature, we get
\eqn\hagid{\beta_H = 4\pi\sqrt{2\ap}/\hbar  =
4\pi/\hbar\oz.}
  That gives us an expression for $b$ upon using
\frcut, \thermII\ and \fail,
\eqn\exb{{bM_0} = \beta_H \qquad\Rightarrow\qquad b = {8\pi G\over \hbar}.}
Putting this back into the equations \thermII\ and \modent, we have
the expressions for the inverse temperature and entropy of our model,
\eqn\same{\beta = {8\pi GM\over \hbar},\qquad
 S = {4\pi GM^2\over \hbar} + 4\pi. }
These expressions are identical to similar expressions for a \bh, equation
\htemp, except for a small difference in the entropy that has a negligible
contribution for large, cold \bh s.

Recently there has been a lot of work aimed at relating the spectrum of
fundamental strings to black hole properties \suss,\sen.  In this letter we
have demonstrated an interesting equality between the temperature and
entropy of a circular string of radius $R$ containing small oscillations,
and a black hole of mass $M=R/2G$. We have presented two distinct results:
the first, equation \thermII, demonstrates the inverse proportionality
between temperature and mass under the plausible assumptions that there
exists an upper frequency cut-off in the oscillations of the string and
that we are considering large cold systems. This corresponds to a slow
collapse of the loop as seen by an observer at infinity. The other result,
equation \same, actually reproduces the Hawking-Bekenstein results. Here
the assumption is that an upper temperature cut-off exists and corresponds
to the Hagedorn temperature in 26 dimensions.  This is also plausible as it
corresponds to the breakdown of the canonical ensemble we have been using,
which diverges above this temperature \ref\turok{N.  Turok, {\sl Physica}
{\bf A158} (1989) 516.}.

It would be interesting to see whether a similar relationship exists for
the case of charged or higher dimensional black holes.

\centerline{\bf Acknowledgements}

We are grateful A. Albrecht, I. Egusquiza, A. Everett, J. Garcia-Bellido,
J.  Garriga, M. Hindmarsh, A. Larsen and A. Liddle for very useful
discussions. A.L.  is supported by PPARC.
\listrefs
\bye